\documentstyle[prl,aps,multicol,psfig]{revtex}
\begin{document}
\draft
\title{Shifting the quantum Hall plateau level in a double layer
electron system}
\author{E.~V. Deviatov, A.~A. Shashkin, and V.~T. Dolgopolov}
\address{Institute of Solid State Physics, Chernogolovka, Moscow
District 142432, Russia}
\author{H.-J. Kutschera and A. Wixforth}
\address{Ludwig-Maximilians-Universit\"at, Geschwister-Scholl-Platz
1, D-80539 M\"unchen, Germany}
\author{K.~L. Campman and A.~C. Gossard}
\address{Materials Department and Center for Quantized Electronic
Structures, University of California, Santa Barbara, California
93106, USA}
\maketitle
\begin{abstract}
We study the plateaux of the integer quantum Hall resistance in a
bilayer electron system in tilted magnetic fields. In a narrow range
of tilt angles and at certain magnetic fields, the plateau level
deviates appreciably from the quantized value with no dissipative
transport emerging. A qualitative account of the effect is given in
terms of decoupling of the edge states corresponding to different
electron layers/Landau levels.
\end{abstract}
\pacs{PACS numbers: 72.20 My, 73.40 Kp}
\begin{multicols}{2}

Much interest in double-layer electron systems is attracted by the
presence of an additional degree of freedom which is associated with
the third dimension. Strong interlayer correlations give rise to the
appearance of novel states that are not observed in single-layer
systems: (i) the even-denominator fractional quantum Hall effect
\cite{eisen,tsui,suen}, (ii) the many-body integer quantum Hall
effect \cite{murphy,lay}, and (iii) broken-symmetry states
\cite{mano}. All the states manifest themselves as quantum plateaux
in the Hall resistance, $\rho_{xy}$, accompanied by zeroes in the
longitudinal resistivity, $\rho_{xx}$. Driving the system out of the
dissipationless regime leads to deviations of $\rho_{xy}$ from the
quantized value, i.e., the behaviour of $\rho_{xy}$ is correlated
with that of $\rho_{xx}$. A deviation of the quantum Hall plateau at
filling factor $\nu=3/2$ from the quantized value accompanied by
non-zero $\rho_{xx}$ was observed in a bilayer system with asymmetric
hole density distributions \cite{hamilton}. Peaks at the low-field
edge (so-called overshoots) of the quantum Hall plateaux along with
corresponding peaks in $\rho_{xx}$ were observed in wide parabolic
GaAs quantum wells in the two-subband regime \cite{ensslin}. Similar
overshoots were previously reported in GaAs/AlGaAs heterostructures
with one occupied subband and explained in terms of
decoupling/depopulation of the edge state associated with the topmost
Landau level \cite{richter}. Normally, additional features on the
quantum Hall plateau are comparable to corresponding ones in
$\rho_{xx}$. On the other hand, whether or not the accuracy of the
Hall resistance quantization is related to dissipative effects solely
is not clear so far. In principle, decoupling of the edge states can
lead to shifting the plateau level in the absence of dissipative
transport as well. In the simplest case of a double layer electron
system with two layers being in different quantum Hall states, the
decoupling of the edge states belonging to different layers can be
easily controlled, e.g., by the application of an in-plane magnetic
field \cite{ensslin,renn}.

Here, we perform precision measurements of the quantized Hall
resistance at integer filling factor in a double layer electron
system in tilted magnetic fields. In a narrow region of tilt angles
and at certain magnetic fields, we observe pronounced deviations of
the quantum Hall plateau from the quantized value which are not
accompanied by any additional features in the dissipative
resistivity. The obtained results are qualitatively explained by
decoupling of the edge states corresponding to different electron
layers/Landau levels, although the sensitivity of the effect to both
tilt angle and magnetic field is unclear.

The samples are grown by molecular beam epitaxy on semi-insulating
GaAs substrate. The active layers form a 760~\AA\ wide parabolic
well. In the center of the well a 3 monolayer thick
Al$_x$Ga$_{1-x}$As ($x=0.3$) sheet is grown which serves as a tunnel
barrier between both parts on either side. The symmetrically doped
well is capped by 600~\AA\ AlGaAs and 40~\AA\ GaAs layers. The
samples are $450\times 50$~$\mu$m$^2$ Hall bars that have a metallic
gate on the crystal surface and ohmic contacts connected to both
electron systems in two parts of the well. Applying a dc voltage
between the well and the gate enables us to tune the carrier density
in the well. The sample is placed in the mixing chamber of a dilution
refrigerator with a base temperature of about 30~mK so that the
normal to its surface is tilted with respect to the magnetic field.
The longitudinal and Hall resistivities of the bilayer electron
system are measured as a function of either magnetic field, $B$, or
gate voltage, $V_g$, using a standard four-terminal lock-in technique
at a frequency of 10~Hz. The excitation current is kept low enough to
ensure that measurements are taken in the linear regime of response.
The data are well reproducible in different coolings of the sample.

For additional magnetocapacitance measurements, a small ac voltage
2.4~mV at frequencies in the range 3 --
\vbox{
\vspace{20mm}
\hbox{
\hspace{-0.2in}
\psfig{file=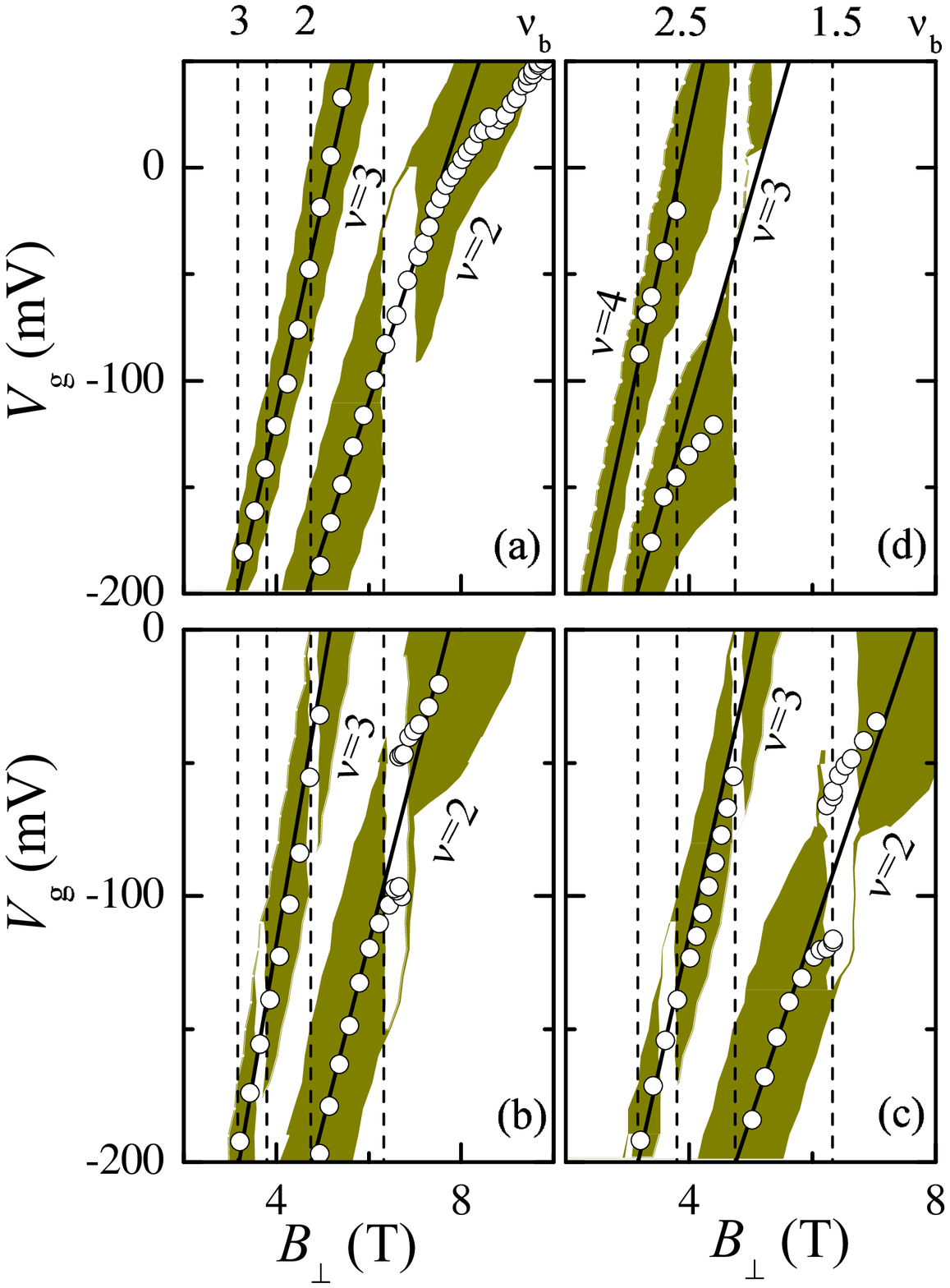,width=3.2in,bbllx=.5in,bblly=1.25in,bburx=7.25in,bbury=9.5in,angle=0}
}
\vspace{-.6in}
\hbox{
\hspace{-0.15in}
\refstepcounter{figure}
\parbox[b]{3.4in}{\baselineskip=12pt \egtrm FIG.~\thefigure.
Positions of the magnetocapacitance minima (circles) in the
($B_\perp,V_g$) plane for $\nu=2,3$, and 4 at tilt angles 45$^\circ$
(a), 50$^\circ$ (b), 53$^\circ$ (c), and 66.5$^\circ$ (d). The dashed
lines correspond to the indicated values of filling factor $\nu_b$ in
the back electron layer. In the shaded areas the deviation of
$\rho_{xy}$ from the quantized value does not exceed
0.05\%.\vspace{0.20in}
}
\label{fig1}
}
}
600~Hz is applied between the
well and the gate and both current components are measured. In the
low frequency limit, the imaginary current component reflects the
thermodynamic density of states in a double layer electron system
whereas the active component of the current is inversely proportional
to the dissipative conductivity (for details, see Ref.~\cite{dqw}).

The positions of the magnetocapacitance minima in the ($B_\perp,V_g$)
plane for filling factor $\nu=2,3$, and 4 are shown in
Fig.~\ref{fig1} by circles for different tilt angles, $\Theta$, of
the magnetic field. Another fan chart (not shown in the figure) is
determined by magnetocapacitance minima corresponding to gaps in the
spectrum of the front electron layer only; these two fan charts allow
determination of the front layer depopulation voltage $V_g=-200$~mV
(bilayer onset) and the voltage $V_g=100$~mV at which the quantum
well becomes symmetric (balance point) \cite{dqw,tilt}. As seen from
Fig.~\ref{fig1}, discontinuities on the fan chart lines for $\nu=2$
and $\nu=3$ emerge with increasing tilt angle. This behavior is
identical with reported earlier on similar samples with the same
quantum well design and interpreted for $\nu=2$ in terms of a
formation of the canted antiferromagnetic phase \cite{caf}.

\vbox{
\vspace{18mm}
\hbox{
\hspace{-0.2in}
\psfig{file=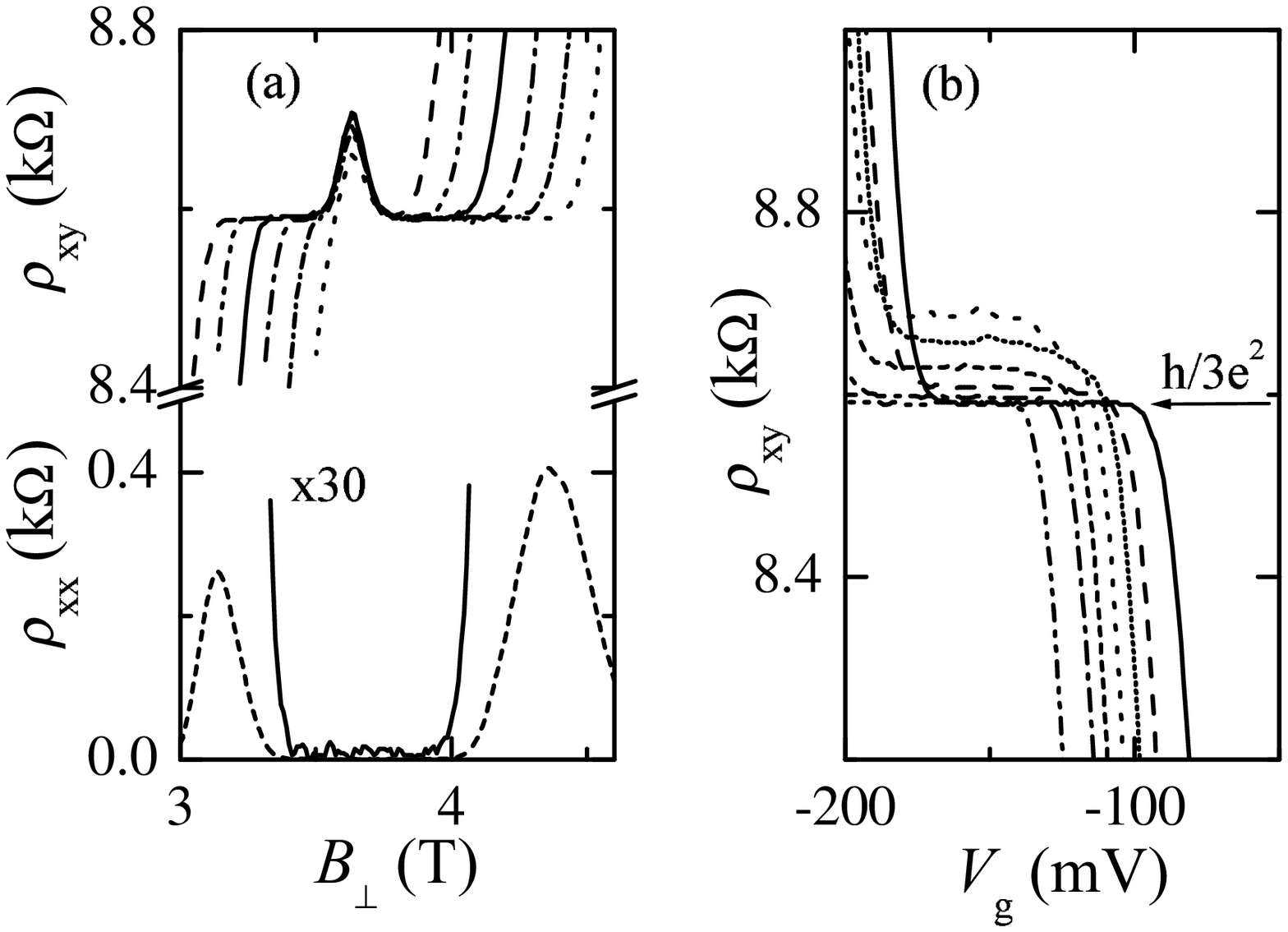,width=3.in,bbllx=.5in,bblly=1.25in,bburx=7.25in,bbury=9.5in,angle=0}
}
\vspace{-1.8in}
\hbox{
\hspace{-0.15in}
\refstepcounter{figure}
\parbox[b]{3.4in}{\baselineskip=12pt \egtrm FIG.~\thefigure.
Traces of $\rho_{xy}$ for $\nu=3$ at $\Theta=53^\circ$ as a function
of $B_\perp$ at gate voltages $-170,-160,-150,-140,-130$, and
$-120$~mV (a) and as a function of $V_g$ at perpendicular components
of the magnetic field $3.45,3.55,3.60,3.65,3.70,3.75$, and 3.85~T
(b). Also shown in case (a) is the dependence of $\rho_{xx}$ on
$B_\perp$ at $V_g=-150$~mV.\vspace{0.20in}
}
\label{fig2}
}
}

In Fig.~\ref{fig2}(a), we show the Hall resistance $\rho_{xy}$ as a
function of magnetic field around filling factor $\nu=3$ for gate
voltages between $-170$ and $-120$~mV at a tilt angle
$\Theta=53^\circ$. There exists a pronounced peak on the quantum Hall
plateau although the longitudinal resistivity $\rho_{xx}$ zeroes
nicely. (We have checked with the help of the magnetocapacitance
measurements that the dissipative conductivity shows no additional
features either.) This peak on the plateau is not sensitive to a
variation of $V_g$ so that at a fixed magnetic field within the peak,
the dependence of $\rho_{xy}$ on gate voltage has a plateau at a
level above the quantized value, see Fig.~\ref{fig2}(b). Such a
behavior of the $\nu=3$ quantum Hall plateau is observed in a narrow
range of tilt angles: it is present for $\Theta=50^\circ$ and
$\Theta=53^\circ$ while at $\Theta=45^\circ$ and $\Theta=66.5^\circ$
the $\nu=3$ quantum Hall plateau is found to be featureless.

Similar shifts of the quantum Hall plateau level accompanied by good
zeroes in $\rho_{xx}$ are also observed at filling factor $\nu=2$ for
$\Theta=45^\circ$, see Fig.~\ref{fig3}(a). Besides, near the
splitting of the $\nu=2$ fan chart line that arises with increasing
$\Theta$ (Fig.~\ref{fig1}), an additional peak appears in both
$\rho_{xx}$ and $\rho_{xy}$ on the shifted plateau, see
Fig.~\ref{fig3}(b). In this case the appearance of a dissipative
transport is naturally reflected by $\rho_{xy}$ behavior.

The overall data on deviation of the quantum Hall plateaux from the
quantized values are depicted in Fig.~\ref{fig1}. The regions in
which the plateau deviation does not exceed 0.05\% are hatched. To
our surprise, at some tilt angles these regions for the same $\nu$
are separated forming regular vertical strips whose position
corresponds to integer and half-integer filling factor, $\nu_b$, in
the back electron layer. Note that the electron density in this layer
is prac-
\vbox{
\vspace{18mm}
\hbox{
\hspace{-0.2in}
\psfig{file=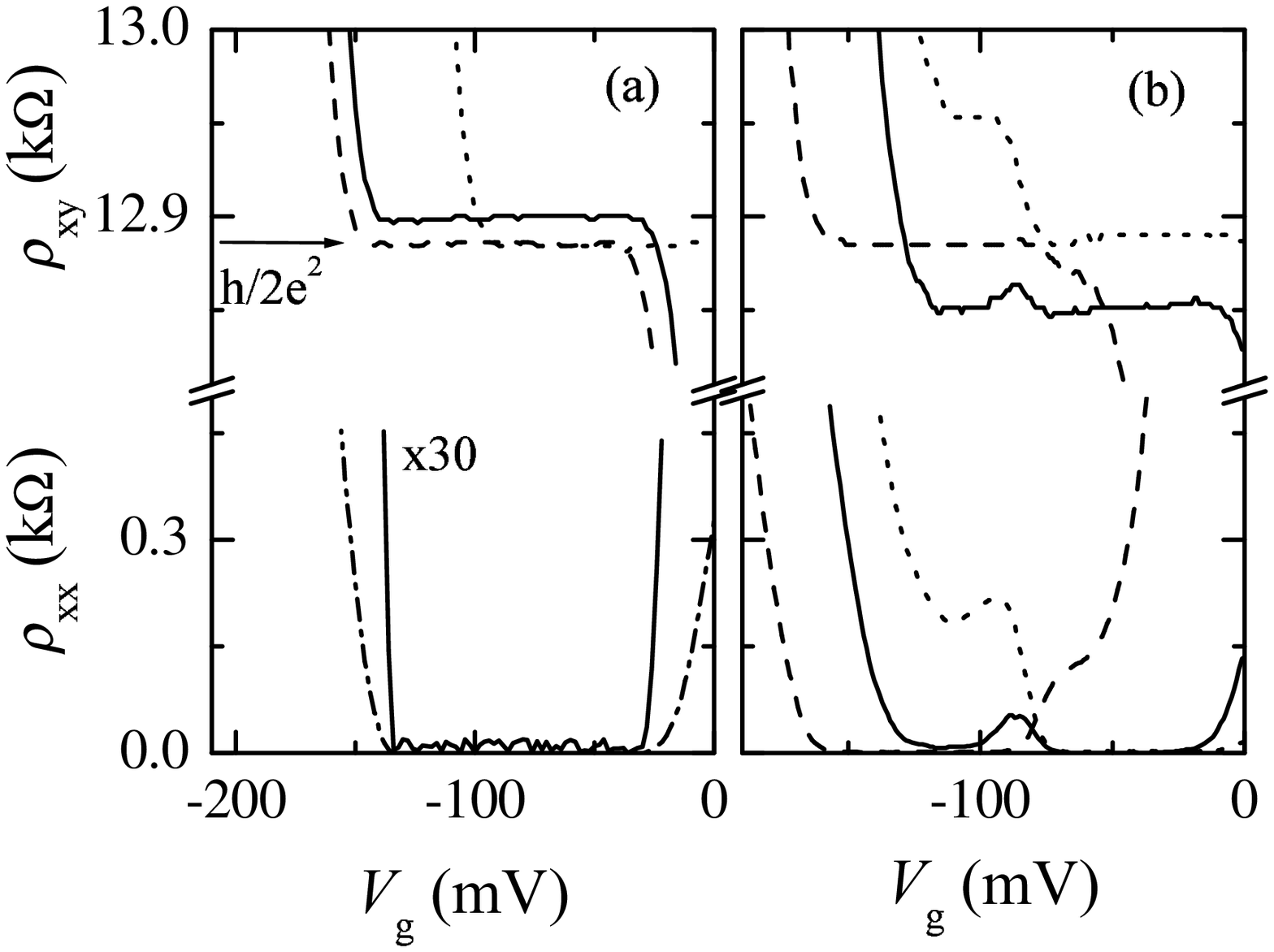,width=3.in,bbllx=.5in,bblly=1.25in,bburx=7.25in,bbury=9.5in,angle=0}
}
\vspace{-1.7in}
\hbox{
\hspace{-0.15in}
\refstepcounter{figure}
\parbox[b]{3.4in}{\baselineskip=12pt \egtrm FIG.~\thefigure.
Dependence of $\rho_{xy}$ and $\rho_{xx}$ on gate voltage for $\nu=2$
at $\Theta=45^\circ$ for $B_\perp=6.26,6.38$, and 7.09~T and for
$B_\perp=6.50$~T, respectively (a) and at $\Theta=53^\circ$ for
$B_\perp=6.04,6.54$, and 6.74~T (b).\vspace{0.20in}
}
\label{fig3}
}
}
tically unchanged with changing $V_g$ because of screening
effect of the front electron layer.

In principle, one can expect possible shifts of the quantum Hall
plateau level: at integer both $\nu_b$ and $\nu$ in an unbalanced
bilayer electron system, two electron layers correspond to two lowest
electron subbands with independent gaps in their spectrum at the
Fermi level, i.e., the electron layers are independent. Therefore,
the condition of inverse proportionality of their individual currents
to $\rho_{xy}$ can be broken, e.g., due to contact resistance,
leading to distinct (decoupled) electrochemical potentials of the
electron layers. Provided that the electrochemical potentials do not
get equilibrated along the edge of the sample including contact
regions, the measured Hall resistance plateau can be above or below
the quantized value even in the absence of dissipative transport.
Similar arguments apply for the observed dissipationless states at
non-integer $\nu_b$ and integer $\nu$. In these states the electron
subbands are correlated as caused by wave function reconstruction in
the unbalanced bilayer electron system \cite{dqw,tilt}. Subject to
the absence of electrochemical potential equilibration between
different Landau levels, the measured Hall resistance plateau can
also be shifted.

However, from the above argumentation it is not clear why deviations
of the quantum Hall plateaux are observed in narrow intervals of the
magnetic field that correspond to integer and half-integer filling
factor $\nu_b$. The sensitivity of the effect to the tilt angle of
magnetic field cannot be explained either. Thus, a more sophisticated
interpretation of the experimental data is needed.

In summary, we have studied the behavior of the quantum Hall
resistance plateaux at integer filling factor in a bilayer electron
system in tilted magnetic fields. In a narrow range of tilt angles
and at magnetic fields corresponding to integer and half-integer
filling factor $\nu_b$, pronounced deviations of the quantum Hall
plateau from the quantized value are observed which are not caused by
dissipative transport. We give a qualitative account of the effect in
terms of decoupling of the edge channels belonging to different
electron layers/Landau levels, although its sensitivity to both tilt
angle and magnetic field is unclear.

We gratefully acknowledge discussions with J.~P. Kotthaus. This work
was supported by the Deutsche Forschungsgemeinschaft under SFB grant
348, RFBR grants 01-02-16424 and 00-02-17294, and INTAS YSF002. The
Munich-Santa Barbara collaboration has also been supported by a joint
NSF-European grant and the Max-Planck research award.




\end{multicols}

\begin{references}
\bibitem{eisen} J.~P. Eisenstein, G.~S. Boebinger, L.~N. Pfeiffer,
K.~W. West, and S. He, Phys.\ Rev.\ Lett.\ {\bf 68}, 1383 (1992).
\bibitem{tsui} Y.~W. Suen, L.~W. Engel, M.~B. Santos, M. Shayegan,
and D.~C. Tsui, Phys.\ Rev.\ Lett.\ {\bf 68}, 1379 (1992).
\bibitem{suen} Y.~W. Suen, H.~C. Manoharan, X. Ying, M.~B. Santos,
and M. Shayegan, Phys.\ Rev.\ Lett.\ {\bf 72}, 3405 (1994).
\bibitem{murphy} S.~Q. Murphy, J.~P. Eisenstein, G.~S. Boebinger,
L.~N. Pfeiffer, and K.~W. West, Phys.\ Rev.\ Lett.\ {\bf 72}, 728
(1994).
\bibitem{lay} T.~S. Lay, Y.~W. Suen, H.~C. Manoharan, X. Ying, M.~B.
Santos, and M. Shayegan, Phys.\ Rev.\ B\ {\bf 50}, 17725 (1994).
\bibitem{mano} H.~C. Manoharan, Y.~W. Suen, T.~C. Lay, M.~B. Santos,
and M. Shayegan, Phys.\ Rev.\ Lett.\ {\bf 79}, 2722 (1997).
\bibitem{hamilton} A.~R. Hamilton, M.~Y. Simmons, F.~M. Bolton, N.~K.
Patel, I.~S. Millard, J.~T. Nicholls, D.~A. Ritchie, and M. Pepper,
Phys.\ Rev.\ B\ {\bf 54}, R5259 (1996).
\bibitem{ensslin} K. Ensslin, A. Wixforth, M. Sundaram, P.~F.
Hopkins, J.~H. English, and A.~C. Gossard, Phys.\ Rev.\ B\ {\bf 47},
1366 (1993).
\bibitem{richter} C.~A. Richter, R.~G. Wheeler, and R.~N. Sacks,
Surf.\ Sci.\ {\bf 263}, 270 (1992).
\bibitem{renn} S.~R. Renn, Phys.\ Rev.\ B\ {\bf 52}, 4700 (1995).
\bibitem{dqw} V.~T. Dolgopolov, A.~A. Shashkin, E.~V. Deviatov, F.
Hastreiter, M. Hartung, A. Wixforth, K.~L. Campman, and A.~C.
Gossard, Phys.\ Rev.\ B\ {\bf 59}, 13235 (1999).
\bibitem{tilt} E.~V. Deviatov, V.~S. Khrapai, A.~A. Shashkin, V.~T.
Dolgopolov, F. Hastreiter, A. Wixforth, K.~L. Campman, and A.~C.
Gossard, JETP\ Lett.\ {\bf 71}, 494 (2000).
\bibitem{caf} V.~S. Khrapai, E.~V. Deviatov, A.~A. Shashkin, V.~T.
Dolgopolov, F. Hastreiter, M. Hartung, A. Wixforth, K.~L. Campman,
and A.~C. Gossard, Phys.\ Rev.\ Lett.\ {\bf 84}, 725 (2000).
\end{references}
\end{document}